# Characteristic Length Scale and Dynamics of $\chi^{3/2}$-MOND Cosmology

Donniel Cruz[1,*] and Emmanuel Rodulfo [2]

[1, 2] *De La Salle University - Manila*
*Corresponding Author: donniel_c_cruz@dlsu.edu.ph / donniel_cruz@dlsu.edu.ph*

**Abstract:** This work studies the cosmology of $\chi^{3/2}$-MOND gravity by Bernal et. al. (2011). This theory is a modification to Einstein's General Relativity (GR) that uses a dimensionless curvature scalar $\chi$ by rescaling the Ricci scalar $R$ by some characteristic length scale $L_M$, as well as a set of modified field equations that follows from a 3/2-power Lagrangian. The characteristic length scale is assumed to be built from the universal constants of the theory and the parameters of the system in question. In the weak field limit, this theory recovers Milgrom's (1983a) Modified Newtonian Dynamics (MOND). MOND is a proposal that corrects Newtonian gravitational laws below an acceleration threshold $a_0 \approx 1.2 \times 10^{-10} m/s^2$ to explain the anomalous flattening of galactic rotation curves without imposing any dark matter components. In the cosmological case, this work asserts that the characteristic length scale is of the order $c^2/a_0$. This specific value is motivated in two ways: (1) it is shown that this scale defines a convergence of GR and MOND at some critical mass (with this as the corresponding length); (2) this length scale is shown to be an extremal value of $L_M$ independent of the mass parameter. The established length scale is then used in the case of cosmology; the FLRW metric is plugged in into the modified field equations and the two modified Friedmann equations are derived incorporating the MOND effects by a manifest appearance of the constant $a_0$.

## 1. INTRODUCTION

Modified Newtonian Dynamics (MOND) is a modification of the laws of (classical, non-relativistic) gravity first proposed by Milgrom (1983a, 1983c, 1983b) in order to explain the anomalous flat rotation curves of galaxies. The standard explanation is that Einstein's General Relativity (GR) is a complete description of (classical) gravity and there must be some extra (invisible) "dark matter" in galaxies that explains the discrepancy in the measured gravitational strength (Rubin & Ford, 1970; Zwicky, 1933). But so far, no such dark matter particle has been directly detected. MOND asserts the opposite; no dark matter particles are needed but instead the laws of gravity must be modified for very weak gravitational strengths.

This proposed correction can be stated as a modification of the Poisson equation for the gravitational scalar field ϕ due so some matter density ρ:

$$\nabla \cdot \left( \mu\left(\frac{|\nabla\phi|}{a_0}\right) \nabla\phi \right) = 4\pi G \rho \qquad (1)$$

where $a_0 \approx 1.2 \times 10^{-10} m/s^2$ is the threshold gravitational acceleration below which MOND effects manifest (Milgrom, 2015); and $\mu(x)$ is some

interpolation function with the following limiting behavior:

$$\mu(x) = \begin{cases} x & \text{for } x \ll 1 \\ 1 & \text{for } x \gg 1 \end{cases} \quad (2)$$

Notice that Eq.(2) reduce to the standard Newtonian gravity in the case when $|\nabla\phi| \gg a_0$.

Around a point mass $M$ in the deep MOND regime (i.e. when $|\nabla\phi| \ll a_0$), the gravitational acceleration is given by:

$$\vec{a} = -\frac{(a_0 GM)^{1/2}}{r}\hat{r} \quad (3)$$

Objects orbiting due to such a gravitational field will have a constant orbital speed $v_{orbital} = \sqrt{|\vec{a}|r} = (a_0 GM)^{1/4}$. This result agrees with the famous baryonic Tully-Fisher relation (McGaugh et al., 2000; Tully & Fisher, 1977), thus giving an explanation to the flattening of the rotation curves in galaxies.

MOND, as it is originally formulated, is not a relativistic theory (it is not Lorentz invariant, etc.). Many relativistic extensions have been proposed, arguably the most famous of which was Tensor-Vector-Scalar (TeVeS) gravity proposed by Bekenstein (2005), but it has fallen out of favor due to its superluminal gravitational wave implications among other things. A recent and more successful attempt is the "New Relativistic MOND" (or RelMOND) proposed by Skordis & Złośnik (2021), which agrees with the cosmic microwave background (CMB) power spectrum, something previously unachievable for many MOND theories.

This paper explores yet another relativistic extension to MOND called $f(\chi)$ gravity (Bernal, Capozziello, Hidalgo, et al., 2011; Mendoza et al., 2012; Bernal et al., 2019). It starts with a generalization of the Einstein-Hilbert action of GR:

$$S_\chi = \frac{c^3}{16\pi G L_M^2}\int d^4x \sqrt{-g}\, f(\chi) \quad (4)$$

where $g \equiv det(g_{\mu\nu})$, $L_M$ is some characteristic length scale dependent on the system at hand, $\chi = L_M^2 R$ is a dimensionless (and rescaled) scalar curvature (with $R$ as the standard Ricci scalar), and $f(\chi)$ is some arbitrary function serving as a degree of freedom of this theory. Notice that when $f(\chi) = \chi$ this action recovers the action of GR. This is very similar to $f(R)$ gravity (Buchdahl, 1970; Sotiriou & Faraoni, 2010). Incorporating a matter term and a cosmological term, the field equations are:

$$f'(\chi)\chi_{\mu\nu} - \frac{1}{2}f(\chi)g_{\mu\nu} - L_M^2(\nabla_\mu\nabla_\nu - g_{\mu\nu}\Box)f'(\chi) + \Lambda L_M^2 g_{\mu\nu} = \frac{8\pi G L_M^2}{c^4}T_{\mu\nu} \quad (5)$$

where $f'(\chi) \equiv df(\chi)/d\chi$, $\chi_{\mu\nu} \equiv L_M^2 R_{\mu\nu}$ is a dimensionless tensor curvature (with $R_{\mu\nu}$ being the Ricci tensor), $\Box \equiv \nabla^\mu\nabla_\mu = g^{\mu\nu}\nabla_\mu\nabla_\nu$ is the 4-dimensional covariant d'Alembertian, $\Lambda$ is a cosmological constant, and $T_{\mu\nu} = -\delta(\sqrt{-g}\mathcal{L}_M)/\delta(2cg^{\mu\nu})$ is the stress-energy tensor (with $\mathcal{L}_M$ being the matter Lagrangian).

For a given gravitational system, the characteristic length scale $L_M$ must be formed from the system's parameters as well as the fundamental constants of the theory. For a point-source mass $M$ (i.e. $T_{00} = M\delta(\vec{x})$, and all other components vanishing), two "fundamental lengths" can be built: the gravitational radius $r_g$ related to GR and the so called "mass-length scale" $l_m$ related to MOND (Bernal, Capozziello, Cristofano, et al., 2011; Milgrom, 2015; Żenczykowski, 2019):

$$r_g \equiv \frac{GM}{c^2} \quad ; \quad l_m \equiv \left(\frac{GM}{a_0}\right)^{1/2} \quad (6)$$

There are many ways to build a "combined" characteristic length scale from these fundamental ones. One such way is to define $L_M$ to be:

$$L_M \equiv \zeta r_g^\alpha l_m^{1-\alpha} \quad (7)$$

where $\zeta$ is some coupling constant of order unity, and $\alpha$ is some real number. Bernal et. al. (2011) showed that the case of $f(\chi) = \chi^{3/2}$ in the field equations Eq. (5) together with the choice $\alpha = ½$ and $\zeta = 2\sqrt{2}/9$ in Eq. (7) and assuming negligible cosmological effects ($\Lambda = 0$) will recover the MOND acceleration law Eq. (3) in the weak-field, slow speed limit.

The gravitational lensing effects (Mendoza et al., 2013) and the dynamics of galaxy clusters (Bernal et al., 2019) in this specific MOND-recovering theory have been thoroughly studied in the literature.

Cosmological extensions of MOND-reducing theories, as well as general modified gravity theories, have been considered by many authors (Clifton et al., 2012; Sotiriou & Faraoni, 2010, and references therein). In this paper, the cosmological implication of this $\chi^{3/2}$-MOND theory is studied. In section 2, the characteristic length scale in the cosmological case is established in two different ways: via "convergence" of MOND and GR, and as an extremal value. In section 3, the modified Friedmann equations are derived by plugging in a cosmological metric tensor ansatz to the modified field equations together with the established length scale from section 2.

## 2. COSMOLOGICAL LENGTH SCALE

The mass-dependent characteristic length scale discussed in §1 is specific to the case of a point mass. In the case of cosmology, one should either consider all the baryonic mass in the universe or one should seek a "critical" length scale independent of mass. Aside from the mass-dependent "fundamental lengths" shown in Eq. (6), there is another mass-independent length that can be built from the constants of this $\chi^{3/2}$-MOND theory:

$$\lambda_M \equiv \frac{c^2}{a_0} \tag{8}$$

The numerical value of this ($\sim 7.39 \times 10^{26}$ m) turns out to be of the same order of magnitude as the Hubble radius (up to a factor of ~6) (Bernal, Capozziello, Cristofano, et al., 2011; Milgrom, 2015; Żenczykowski, 2019). The following establishes two ways arrive at this length.

### 2.1 Convergence of MOND and GR

The two fundamental lengths in Eq. (6) corresponding to GR and MOND are both functions of mass. Roughly speaking, for a fixed point-source $M$, $r_g$ defines the distance in which GR effects dominate (cf. with the Schwarzschild radius), while $l_m$ establishes the region where MOND effects "begin" (the Newtonian gravitational acceleration is exactly equal to $a_0$ at a radial distance of $l_m$ from the source). Solving for the non-zero critical mass $M = M^*$ such that $r_g$ is equal to $l_m$, one gets:

$$r_g \big|_{M=M^*} = l_m \big|_{M=M^*} \implies M^* = \frac{c^4}{a_0 G} \tag{9}$$

The numerical value of this mass ($\sim 1.01 \times 10^{54}$ kg) is of the same order or magnitude as the total baryonic mass of the observable universe. Plugging in this critical mass to the fundamental lengths $r_g^* \equiv r_g|_{M=M^*}$ and $l_m^* \equiv l_m\big|_{M=M^*}$:

$$r_g^* = l_m^* = \frac{c^2}{a_0} = \lambda_M \tag{10}$$

Using the definition in Eq. (7), the cosmological length scale $L_M^*$ is given to be:

$$L_M^* = \zeta \frac{c^2}{a_0} = \zeta \lambda_M \tag{11}$$

This method of arriving at this value is independent of the exponent parameter $\alpha$ that appears in Eq. (7).

### 2.2 Extremal value

One can also consider the extremal values of the characteristic length scale defined in Eq. (7). Taking the derivative of $L_M$ with respect to $M$ and setting this value to zero, one gets:

$$\frac{d}{dM} L_M = \zeta \frac{d}{dM} \left( \left(\frac{GM}{c^2}\right)^\alpha \left(\frac{GM}{a_0}\right)^{\frac{1-\alpha}{2}} \right)$$
$$0 = \zeta \frac{G^{(\alpha+1)/2}}{c^{2\alpha} a_0^{(1-\alpha)/2}} \left(\frac{\alpha+1}{2}\right) M^{(\alpha-1)/2} \tag{12}$$

Assuming that $M \neq 0$, the only way for the Eq. (12) to be satisfied is when $\alpha = -1$. This implies that the cosmological characteristic length scale $L_M^*$ is:

$$\begin{aligned} L_M^* &= L_M \big|_{\alpha=-1} \\ &= \zeta \left(\frac{GM}{c^2}\right)^{-1} \left(\frac{GM}{a_0}\right)^1 \\ &= \zeta \frac{c^2}{a_0} \end{aligned} \tag{13}$$

which is identical to the result Eq. (11) in §2.1. This method implies that an extremal value is only possible when the exponent parameter $\alpha = -1$ and $L_M$ is a constant quantity independent of mass.

## 3. METRIC TENSOR, SCALE FACTOR, AND COSMOLOGY

The expansion of the universe can be modeled in GR using the Friedmann-Lemaître-Robertson-Walker (FLRW) metric which describes a homogenous and isotropic universe whose spatial scale is allowed to evolve in time. The line element in spherical polar coordinates and with $(-, +, +, +)$ signature is:

$$\begin{aligned} ds^2 = -c^2 d\tau^2 = &-c^2 dt^2 \\ &+ a(t)^2 \left( \frac{1}{1-kr^2} dr^2 \right. \\ &\left. + r^2 d\theta^2 + r^2 \sin^2 \theta \, d\phi^2 \right) \end{aligned} \tag{14}$$

where $a(t)$ is a time-dependent scale factor, and $k$ is a constant which can take values $-1$, $0$, or $+1$, corresponding to a closed, flat, or open universe respectively (Carroll, 2004; Wald, 1984).

Using the FLRW metric in Eq. (14) and the characteristic length scale established in §2, the dimensionless scalar curvature $\chi$ evaluates to be:

$$\chi = \frac{6\zeta^2 \lambda_M^2}{c^2} \frac{(a\ddot{a} + \dot{a}^2 + kc^2)}{a^2} \tag{15}$$

where an overdot denotes differentiation with respect to $t$, i.e. $\dot{a} = da/dt$ and $\ddot{a} = d^2a/dt^2$. Similarly, the non-zero (and incidentally diagonal) components of the dimensionless tensor curvature $\chi_{\mu\nu}$ are:

$$\chi_{00} = -\frac{3\zeta^2 \lambda_M^2}{c^2}\frac{\ddot{a}}{a} \quad (16)$$

$$\chi_{11} = \frac{\zeta^2 \lambda_M^2}{c^2}\frac{a\ddot{a} + 2\dot{a}^2 + 2kc^2}{1 - kr^2} \quad (17)$$

$$\chi_{22} = \frac{\zeta^2 \lambda_M^2}{c^2}(a\ddot{a} + 2\dot{a}^2 + 2kc^2)r^2 \quad (18)$$

$$\chi_{33} = \chi_{22} \sin^2\theta \quad (19)$$

The matter content of a homogenous and isotropic universe can be modeled by a perfect fluid with mass density $\rho$, pressure $p$, and dimensionless 4-velocity $U$. The corresponding energy-momentum tensor in the comoving frame is $T_{\mu\nu} = (\rho c^2 + p)U_\mu U_\nu + p g_{\mu\nu}$, or with mixed indices is $T^\mu{}_\nu = \text{diag}[-\rho c^2, p, p, p]$; its trace is $T \equiv g^{\mu\nu}T_{\mu\nu} = -\rho c^2 + 3p$ (Carroll, 2004). One can also define a the "dark energy density" as $\rho_\Lambda = \Lambda c^2/8\pi G$. By plugging in this information into the Einstein field equations, one can derive the Friedmann equations:

$$\frac{\ddot{a}}{a} = -\frac{4\pi G}{3}\left(\rho + \frac{3}{c^2}p - 2\rho_\Lambda\right) \quad (20)$$

$$\frac{\dot{a}^2 + kc^2}{a^2} = \frac{8\pi G}{3}(\rho + \rho_\Lambda) \quad (21)$$

The Friedmann equations Eq. (20) and Eq. (21) imply that $R > 4\Lambda$ for a non-empty universe (i.e., there is matter and/or radiation together with the cosmological constant). The MOND acceleration threshold $a_0$ numerically turns out to be (possibly only coincidentally) related to the cosmological constant $\Lambda$, in particular $2\pi a_0 \approx c^2\sqrt{\Lambda/3}$ (Bernal, Capozziello, Cristofano, et al., 2011; Milgrom, 2015; Żenczykowski, 2019). Whether or not this is pure coincidence or is a derivable relation from theory is beyond the scope of this paper. This does imply that $\chi \gg 1$ as long as the characteristic length scale coupling parameter $\zeta$ is of order unity. This magnitude for $\chi$ shall be assumed in the case of $\chi^{3/2}$-MOND cosmology. The field equations for the case of $f(\chi) = \chi^{3/2}$ can then be written as:

$$\frac{3}{2}\chi^{1/2}\chi_{\mu\nu} - \frac{1}{2}\chi^{3/2}g_{\mu\nu} + \mathcal{O}(\chi^{1/2})$$
$$= \frac{8\pi G}{c^4}L_M^{*2}T_{\mu\nu} - \Lambda L_M^{*2}g_{\mu\nu} \quad (22)$$

And since $\chi^{1/2} \ll \chi^{3/2}$, the higher derivative terms can be ignored in this approximate treatment. This also makes the field equations "Friedmann-like" in the sense that only terms like $\ddot{a}$ and $\dot{a}^2$ appear explicitly in the field equations; the full field equations with the $\mathcal{O}(\chi^{1/2})$-terms kept contain terms like $\dddot{a}, \ddot{a}\dot{a}, \dot{a}^3, \ddddot{a}$, and so on.

Considering only the "00" term of the field equations and plugging in the scalar curvature from Eq. (15) and the first component of the tensor curvature from Eq. (16), one obtains the following relation:

$$\zeta\sqrt{6}\frac{c}{a_0}\left(\frac{\dot{a}^2 + kc^2}{a^2} + \frac{\ddot{a}}{a}\right)^{1/2}\left(\frac{\dot{a}^2 + kc^2}{a^2} - \frac{1}{2}\frac{\ddot{a}}{a}\right)$$
$$= \frac{8\pi G}{3}(\rho + \rho_\Lambda) \quad (23)$$

And taking the trace of the field equations one gets:

$$\zeta\frac{\sqrt{6}}{2}\frac{c}{a_0}\left(\frac{\dot{a}^2 + kc^2}{a^2} + \frac{\ddot{a}}{a}\right)^{3/2}$$
$$= \frac{4\pi G}{3}\left(\rho - \frac{3}{c^2}p + 4\rho_\Lambda\right) \quad (24)$$

Equations Eq. (23) and (24) can be regarded as a system of two algebraic equations in terms of the variables $\ddot{a}/a$ and $(\dot{a}^2 + kc^2)/a^2$. Assuming all parameters are real, one can exactly solve this system to arrive at the following modified Friedmann equations:

$$\frac{\ddot{a}}{a} = -\frac{4\sqrt[3]{4}}{3}\left(\frac{\pi G a_0}{\zeta c}\right)^{2/3}\frac{\frac{1}{c^2}p - \rho_\Lambda}{\left(\rho - \frac{3}{c^2}p + 4\rho_\Lambda\right)^{1/3}} \quad (25)$$

$$\frac{\dot{a}^2 + kc^2}{a^2} = \frac{2\sqrt[3]{4}}{3}\left(\frac{\pi G a_0}{\zeta c}\right)^{2/3}\frac{\rho - \frac{1}{c^2}p + 2\rho_\Lambda}{\left(\rho - \frac{3}{c^2}p + 4\rho_\Lambda\right)^{1/3}} \quad (26)$$

In contrast with the GR case in Eq. (20) and Eq. (21), the modified Friedmann equations Eq. (25) and Eq. (26) are highly non-linear and shall require further investigation. It's worth noting that the denominator term suggests the existence of specific combinations of $\rho$, $p$, and $\Lambda$ that will render these equations singular. Plugging these results into the definition of the dimensionless curvature scalar $\chi$ in Eq. (15) and noting that $2\pi a_0 \approx c^2\sqrt{\Lambda/3}$, one indeed finds $\chi \gg 1$, justifying the assumption made for Eq. (22) and in the passage that follows it above.

## 4. CONCLUSIONS

This work presented two ways to motivate an appropriate characteristic length scale for the cosmological treatment of $\chi^{3/2}$-MOND gravity. The two methods in §2 can be viewed as two ways of finding the extreme values of $L_M$. This derived cosmological length numerically turns out to be of the same order of magnitude as current Hubble radius.

Modified Friedmann equations were also derived by plugging in the FLRW metric to the modified field equations and using the established cosmological length scale in the definition of the dimensionless curvature quantities. It was argued that numerically $\chi \gg 1$, which allows for the omission of the higher derivative terms in the field equations as an approximation. The pair of equations derived are highly non-linear but are manifestly MONDian due to the explicit appearance of Milgrom's acceleration threshold $a_0$.

Further investigation on the behavior of these pair of equations is needed for the various possible energy density contents of the universe (matter, radiation, dark energy, etc.). The singular behavior of the equations also warrants a closer analysis.

Finally, omitting the higher derivative terms in the field equations explicitly violates local momentum conservation (i.e., $\nabla_\mu T^{\mu\nu} = 0$); ways to recover this conservation law while maintaining the second-order nature of the modified Friedmann equations is a topic the authors are working on at the time of writing this paper.

## 5. ACKNOWLEDGMENTS

The author D.Cruz thanks the DOST-ASTHRDP scholarship program for the financial support of his studies, which made this work possible.

## 6. REFERENCES

Bekenstein, J. D. (2005). Relativistic gravitation theory for the MOND paradigm. *Physical Review D*, *71*(6), 069901. https://doi.org/10.1103/PhysRevD.71.069901

Bernal, T., Capozziello, S., Cristofano, G., & De Laurentis, M. (2011). MOND'S ACCELERATION SCALE AS A FUNDAMENTAL QUANTITY. *Modern Physics Letters A*, *26*(36), 2677–2687. https://doi.org/10.1142/S0217732311037042

Bernal, T., Capozziello, S., Hidalgo, J. C., & Mendoza, S. (2011). Recovering MOND from extended metric theories of gravity. *The European Physical Journal C*, *71*(11), 1794. https://doi.org/10.1140/epjc/s10052-011-1794-z

Bernal, T., López-Corona, O., & Mendoza, S. (2019). DYNAMICS OF CLUSTERS OF GALAXIES WITH EXTENDED F(chi) GRAVITY. *Revista Mexicana de Astronomía y Astrofísica*, *55*(2), 237–254. https://doi.org/10.22201/ia.01851101p.2019.55.02.12

Buchdahl, H. A. (1970). Non-Linear Lagrangians and Cosmological Theory. *Monthly Notices of the Royal Astronomical Society*, *150*(1), 1–8. https://doi.org/10.1093/mnras/150.1.1

Carroll, S. M. (2004). *Spacetime and geometry: An introduction to general relativity*. Addison Wesley.

Clifton, T., Ferreira, P. G., Padilla, A., & Skordis, C. (2012). Modified Gravity and Cosmology. *Physics Reports*, *513*(1–3), 1–189. https://doi.org/10.1016/j.physrep.2012.01.001

McGaugh, S. S., Schombert, J. M., Bothun, G. D., & de Blok, W. J. G. (2000). The Baryonic Tully-Fisher Relation. *The Astrophysical Journal*, *533*, L99–L102. https://doi.org/10.1086/312628

Mendoza, S., Bernal, T., Hernandez, X., Hidalgo, J. C., & Torres, L. A. (2013). Gravitational lensing with f ($\chi$) = $\chi$3/2 gravity in accordance with astrophysical observations. *Monthly Notices of the Royal Astronomical Society*, *433*(3), 1802–1812. https://doi.org/10.1093/mnras/stt752

Mendoza, S., Bernal, T., Hidalgo, J. C., & Capozziello, S. (2012). *MOND as the weak-field limit of an extended metric theory of gravity*. 483–486. https://doi.org/10.1063/1.4734465

Milgrom, M. (1983a). A modification of the Newtonian dynamics as a possible alternative to the hidden mass hypothesis.

*The Astrophysical Journal*, *270*, 365–370. https://doi.org/10.1086/161130

Milgrom, M. (1983b). A modification of the newtonian dynamics: Implications for galaxy systems. *The Astrophysical Journal*, *270*, 384–389. https://doi.org/10.1086/161132

Milgrom, M. (1983c). A modification of the Newtonian dynamics—Implications for galaxies. *The Astrophysical Journal*, *270*, 371–383. https://doi.org/10.1086/161131

Milgrom, M. (2015). MOND theory. *Canadian Journal of Physics*, *93*(2), 107–118. https://doi.org/10.1139/cjp-2014-0211

Rubin, V. C., & Ford, W. K., Jr. (1970). Rotation of the Andromeda Nebula from a Spectroscopic Survey of Emission Regions. *The Astrophysical Journal*, *159*, 379. https://doi.org/10.1086/150317

Skordis, C., & Złośnik, T. (2021). New Relativistic Theory for Modified Newtonian Dynamics. *Physical Review Letters*, *127*(16), 161302. https://doi.org/10.1103/PhysRevLett.127.161302

Sotiriou, T. P., & Faraoni, V. (2010). F(R) Theories Of Gravity. *Reviews of Modern Physics*, *82*(1), 451–497. https://doi.org/10.1103/RevModPhys.82.451

Tully, R. B., & Fisher, J. R. (1977). A new method of determining distances to galaxies. *Astronomy and Astrophysics*, *54*, 661–673. https://ui.adsabs.harvard.edu/abs/1977A&A....54..661T

Wald, R. M. (1984). *General relativity*. University of Chicago Press.

Żenczykowski, P. (2019). MOND and natural scales of distance and mass. *Modern Physics Letters A*, *34*(37), 1950306. https://doi.org/10.1142/S0217732319503061

Zwicky, F. (1933). Die Rotverschiebung von extragalaktischen Nebeln. *Helvetica Physica Acta*, *6*, 110–127. https://ui.adsabs.harvard.edu/abs/1933AcHPh...6..110Z